\documentclass[conference]{IEEEtran}
\usepackage{caption}
\usepackage{algorithm}
\usepackage{algpseudocode}
\usepackage{cite}
\usepackage[T1]{fontenc}
\usepackage{amsmath,amssymb,amsfonts}
\usepackage{newtxtext,newtxmath}
\usepackage{graphicx}
\usepackage{textcomp}
\usepackage[dvipsnames,table]{xcolor}  % Add 'dvipsnames' to support colors

\usepackage{underscore}
\usepackage[english]{babel}
\usepackage{subfigure}
\usepackage{float} 

\usepackage{enumitem}
\usepackage{titlesec}
\usepackage{balance} 
\usepackage{menukeys}
\usepackage{booktabs}
\def\BibTeX{{\rm B\kern-.05em{\sc i\kern-.025em b}\kern-.08em
    T\kern-.1667em\lower.7ex\hbox{E}\kern-.125emX}}

\makeatletter
\newcommand{\linebreakand}{%
  \end{@IEEEauthorhalign}
  \hfill\mbox{}\par
  \mbox{}\hfill\begin{@IEEEauthorhalign}
}
\makeatother

\begin{document}

\title{OpenCat: Improving Interoperability of ADS Testing}

\author{\IEEEauthorblockN{Qurban Ali}
\IEEEauthorblockA{
\textit{University of Milano-Bicocca}\\
Milan, Italy \\
q.ali@campus.unimib.it}
\and
\IEEEauthorblockN{Andrea Stocco}
\IEEEauthorblockA{
\textit{Technical University of Munich, fortiss GmbH}\\
Munich, Germany \\
andrea.stocco@tum.de, stocco@fortiss.org}

\linebreakand
\hspace{-1.0cm} 
\IEEEauthorblockN{Leonardo Mariani}
\IEEEauthorblockA{
\hspace{-1.4cm}\textit{University of Milano-Bicocca}\\
\hspace{-1.4cm}Milan, Italy \\
\hspace{-1.0cm}leonardo.mariani@unimib.it}
\and
\IEEEauthorblockN{Oliviero Riganelli}
\IEEEauthorblockA{
\textit{University of Milano-Bicocca}\\
Milan, Italy \\
oliviero.riganelli@unimib.it}
}

\maketitle

\begin{abstract}

Testing Advanced Driving Assistance Systems (ADAS), such as lane-keeping functions, requires creating road topologies or using predefined benchmarks. However, the test cases in existing ADAS benchmarks are often designed in specific formats (e.g., OpenDRIVE) and tailored to specific ADAS models. This limits their reusability and interoperability with other simulators and models, making it challenging to assess ADAS functionalities independently of the platform-specific details used to create the test cases.

This paper evaluates the interoperability of \textsc{SensoDat}, a benchmark developed for ADAS regression testing. We introduce \textsc{OpenCat}, a converter that transforms OpenDRIVE test cases into the Catmull-Rom spline format, which is widely supported by many current test generators. By applying \textsc{OpenCat} to the \textsc{SensoDat} dataset, we achieved high accuracy in converting test cases into reusable road scenarios. 
To validate the converted scenarios, we used them to evaluate a lane-keeping ADAS model using the Udacity simulator. Both the simulator and the ADAS model operate independently of the technologies underlying \textsc{SensoDat}, ensuring an unbiased evaluation of the original test cases.
Our findings reveal that benchmarks built with specific ADAS models hinder their effective usage for regression testing. We conclude by offering insights and recommendations to enhance the reusability and transferability of ADAS benchmarks for more extensive applications.
\end{abstract}


\section{Introduction}

Autonomous Driving Systems (ADS) integrate advanced technologies like adaptive cruise control, parking assistance, and autopilots into a cohesive functional unit. Modern ADS are designed with enhanced capabilities to operate independently with minimal or no human intervention, following a perception-plan-execution framework~\cite{survey-lei-ma,yurtsever2020survey}. The perception component is delegated to Advanced Driver Assistance Systems (ADAS) operated by sophisticated perception systems, including cameras, LiDAR, and various sensors, to interpret complex and dynamic driving environments. These perception systems, in turn, leverage deep neural networks (DNNs) to process sensor data and perform tasks such as object detection, image classification, semantic segmentation, and regression, thereby enabling real-time driving functionalities~\cite{survey-lei-ma,yurtsever2020survey,li2024panopticperceptionautonomousdriving,grigorescu2020survey}.

Validating autonomous driving systems across diverse scenarios requires designing extensive testing and simulation environments~\cite{2020-Riccio-EMSE}, such as the creation of detailed road topologies, to test the limitations of lane-keeping components~\cite{Elghazaly2023HighDefinitionMC,banafa2024,2022-Stocco-ASE,2020-Stocco-ICSE,2023-Stocco-EMSE,2021-Stocco-JSEP,2024-Lambertenghi-ICST,2025-Lambertenghi-ICST}, among others. Testing ADAS with challenging road topologies in virtual simulation environments is crucial, not only to drastically reduce the cost of testing autonomous vehicles but also to provide safety guarantees before the software is deployed in production on physical vehicles~\cite{2023-Stocco-TSE,2023-Stocco-EMSE,kim2017testing,zofka2016testing}.
 
Recent years have seen a growing focus on testing methodologies for ADAS/ADS, particularly concerning test generation techniques~\cite{NeelofarTOSEM,NeelofarICSE,pafot,epitester,lu2023deepqtesttestingautonomousdriving,pan2023safetyassessmentvehiclecharacteristics,KLUCK2023107225,10685189,HUMENIUK2023102990,10.1145/3650105.3652296,ARCAINI2024103171,10.1145/3550270,10.1007/978-3-031-49269-3_14,10.1007/978-3-031-49266-2_6,2025-Baresi-ICSE, PDL}. Other studies have proposed benchmarks for ADAS testing, such as \textsc{SensoDat}~\cite{sensodat}, \textsc{DeepScenario}~\cite{deepscenario}, and \textsc{SCTrans}~\cite{sctrans}. These benchmarks include hundreds of scenarios designed for ADAS regression testing, intended to provide reusable specifications for testing. However, instead of defining the test objectives in a general and reusable manner, they often focus on technical implementations that detail specific test executions for particular simulation platforms and ADAS models. These implementations include both successful and failing test cases, which may not adequately represent corner cases across different ADAS models or simulators~\cite{borg,AminiFlaky2024}.
Due to the tight coupling between test implementations, simulation platforms, and ADAS models, we hypothesize that this technical dependency limits the broader applicability and reuse of these benchmarks.

In this paper, we evaluate our preliminary hypothesis by examining the interoperability of high-level scenarios, with a specific focus on \textsc{SensoDat}, and investigate their potential to provide truly reusable specifications. To achieve this, we adapt \textsc{SensoDat} scenarios for use in a different simulator with a distinct road format, testing their compatibility with an independent, learning-based ADAS model.

First, we describe \textsc{OpenCat}, a tool designed to convert the OpenDRIVE road format (used by \textsc{SensoDat} and \textsc{BeamNG.tech}~\cite{beamng}) into the Catmull-Rom spline representation (utilized by the Udacity simulator~\cite{udacity_sim}). Both formats are integral to the field of ADS/ADAS development. OpenDRIVE~\cite{dupuis2010opendrive} is widely adopted in the industry for defining road networks and is commonly used to test \mbox{Level 4} ADS in complex urban environments. In contrast, Catmull-Rom splines~\cite{Catmull1974ACO} offer a smooth and mathematically precise representation of roads, making them ideal for generating controllable road scenarios to test \mbox{Level 2} ADAS systems, such as trajectory planning or lane-keeping assistance.

While \textsc{OpenCat} provides precise conversion from OpenDRIVE data to Catmull-Rom splines (100\% accuracy), our second contribution focuses on evaluating the interoperability of the \textsc{SensoDat} benchmark. Re-simulation experiments conducted in the Udacity self-driving car simulator reveal significant percentage differences in the test pass ratio, ranging from 25\% to 50\%, emphasizing that benchmarks closely tailored to specific ADAS models may constrain their broader applicability for regression testing.
Our paper makes the following contributions: 

\begin{enumerate}
    \item To the best of our knowledge, we present the first openly accessible converter \textsc{OpenCat}, that transforms the OpenDRIVE road format to a Catmull-Rom spline representation.
    \item We show the utility of \textsc{OpenCat} by applying it to the publicly available \textsc{SensoDat}~\cite{sensodat} dataset, successfully converting 32,580 roads with high accuracy.
    \item We validate the practical usability of the converted roads through simulations in the Udacity self-driving car simulator using ADAS available from the literature. Our findings show how reliance on specific ADAS models impacts the interoperability of existing scenarios in \textsc{SensoDat}, thereby calling for solutions to enhance their reusability. 
\end{enumerate}

The paper is organized as follows: Section~\ref{sec:background} provides background information about lane-keeping ADAS and the \textsc{SensoDat} benchmark. Section~\ref{sec:overview} overviews the OpenDRIVE format and the Catmull Rom spline. Section~\ref{sec:conversion} presents \textsc{OpenCat}, that is, the conversion algorithm used for translating roads from the OpenDRIVE to the Catmull Rom Spline format. The results and experiments are discussed in Section ~\ref{sec:results}. Finally, Section~\ref{sec:conclusion} provides concluding remarks.
% ANDREA: this part is old fashioned and no longer used in short conference and workshop papers

\section{Background} \label{sec:background}

\subsection{Lane-keeping ADAS}

We consider NHTSA~\cite{nhtsa} Level 2 ADAS, which relies on vision-based perception tasks using data captured by vehicle-mounted sensors. 
This paper focuses on ADAS which provides the \textit{lane-keeping} functionality from human-labeled driving data. Lane-keeping is a critical component for the safe operation of ADS. 
Models such as NVIDIA's Dave-2~\cite{nvidia-dave2} learn to drive by identifying latent patterns in a training dataset composed of images captured during expert human driving. These models predict driving commands that replicate the behavior of a skilled driver. In its simplest form, a lane-keeping ADAS, such as Dave-2, can be represented as a function $f: \mathbb{R}^d \rightarrow [-25^\circ, +25^\circ]$. Here, $d$ denotes the dimension of the input image $\boldsymbol{x} \in \mathbb{R}^d$ (e.g., for an image of size $140 \times 320$, $d = 44,800$ pixels), and the output is a scalar value $y$ representing the predicted steering angle. This angle lies within the range \big[$-25^\circ, +25^\circ$\big], where $-25^\circ$ indicates maximum steering to the left, $+25^\circ$ indicates maximum steering to the right, and $0^\circ$ corresponds to no steering input.\footnote{These values reflect the steering limits of an ADAS within the Udacity driving simulator~\cite{udacity-simulator}.}

\subsection{Benchmark}

The \textsc{SensoDat} dataset~\cite{sensodat} comprises 32,580 test scenarios (i.e., roads) generated using existing test case generators. These scenarios encompass a wide range of road conditions, including curved roads, sharp turns, and complex layouts, enabling the testing of self-driving systems, such as lane-keeping models. The test scenarios were executed within the BeamNG simulator~\cite{beamng} using its AI driver autopilot,\footnote{This conservative assumption is made due to the lack of specific details provided in the original paper.} a PID-based controller with global knowledge of the road topology. While \textsc{SensoDat} mainly serves as a benchmark for sensor data, the authors highlight its potential for regression testing purposes, as trajectory data are also included. The dataset reports 19,926 successful test executions and 12,654 failures.

In this paper, we explore the interoperability of \textsc{SensoDat} within the context of a different simulator (Udacity) and a learning-based driving model (Dave-2).

\section{Overview of OpenDRIVE format and Catmull Rom spline} \label{sec:overview}

\textsc{SensoDat} uses the OpenDRIVE format, which required us to convert the road topologies to the Catmull-Rom spline format used in Udacity. In this section, we provide a concise overview of the OpenDRIVE road format and Catmull-Rom spline, highlighting the key differences, followed by a detailed explanation of our conversion method in Section~\ref{sec:conversion}. 

\begin{figure}[!tbp]
\centering
\includegraphics[width=0.49\textwidth]{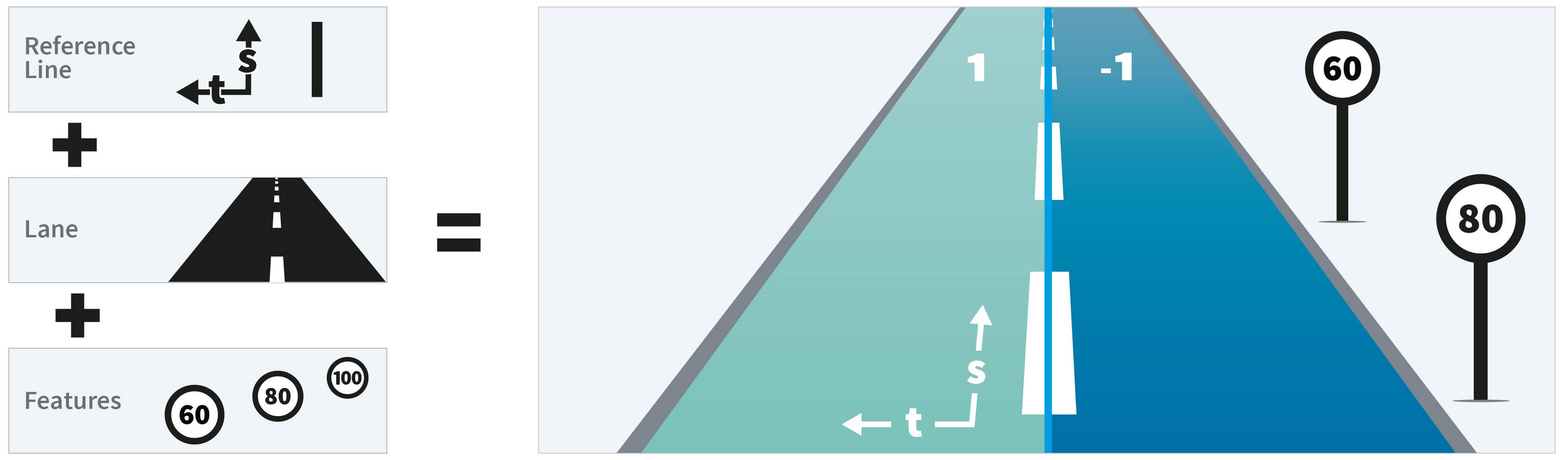}
\caption{\label{fig:rline}Elements of OpenDRIVE~\cite{asam_opendrive}.}
\end{figure}

\subsection{OpenDRIVE}

ASAM OpenDRIVE~\cite{asam_opendrive} provides a comprehensive extensible markup language (XML)-based format (.xodr) for modeling networks in ADS simulators. An OpenDRIVE file contains data that describes the geometry of roads, lanes, and objects---such as road markings---as well as features along the roads, including signals as shown in Figure~\ref{fig:rline}. The reference line, the backbone of the format, is constructed using concatenated clothoids (Euler spirals)~\cite{baass1984clothoid}, arcs, and polynomials. This ensures smooth transitions in road curvature. In addition, lanes are specified by type, width, and position relative to the reference line and may be modified sectionally. Junctions are modeled by linking roads and lanes, allowing for complex traffic simulations~\cite{isprs}. 

\textsc{SensoDat} does not include junctions, as the roads are treated individually rather than as a connected map: each road is identified by a unique road ID and processed separately. Additionally, \textsc{SensoDat} does not exploit the flexibility of OpenDRIVE, and in this case, the richness of the format is mostly a source of complexity. Thus, researchers have proposed alternative road formats~\cite{klikovits2022doesroaddiversityreally}, like the Catmull-Rom spline. Unlike OpenDRIVE, it does not rely on a reference line and provides a simpler and more efficient road representation. 

\begin{figure}[!tbp]
\centering
\includegraphics[width=0.3\textwidth, height=3cm, keepaspectratio]{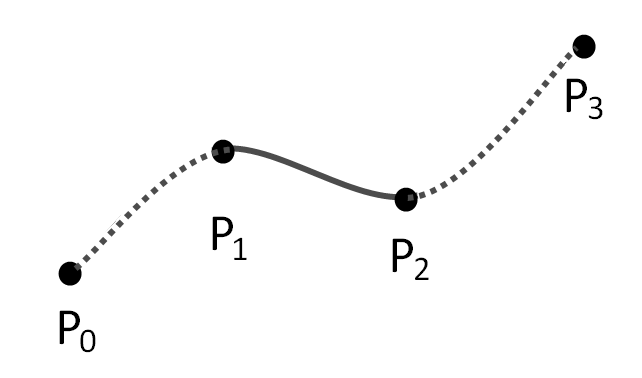}
\caption{\label{fig:catmull}Catmull-Rom spline interpolation with four points.}
\end{figure}
\subsection{Catmull-Rom Spline}

A Catmull-Rom spline~\cite{Catmull1974ACO} is a smooth curve (interpolation spline) widely used in computer graphics, animation, and path generation to create natural transitions between points. It passes directly through each control point (no approximation) while maintaining smoothness, meaning that there are no sharp angles or sudden changes in direction. One of its advantages is its localization control, which means that changing one point only affects the neighboring segments, not the entire curve, making it easy to tweak without disturbing the whole shape. 

From a mathematical point of view, the spline uses cubic polynomials to calculate how the curve behaves between four consecutive points; the key idea here is that it creates a smooth transition between any two points while considering their neighbor for context. There is also a tension parameter that controls how tightly the curve bends around the points. Higher tension leads to sharper turns, while lower tension results in a more relaxed flowing curve~\cite{Catmull1974ACO}. 

Given a series of control points \( P_0, P_1, P_2, P_3 \) (Figure~\ref{fig:catmull}), the Catmull-Rom spline segment between \( P_1 \) and \( P_2 \) is computed using the following cubic Hermite interpolation formula:

\begin{equation}
% \small
\textstyle
\label{eq1}
\begin{aligned}
C(t) = \frac{1}{2} \Big[ & (2P_1) + (-P_0 + P_2)t + (2P_0 - 5P_1 + 4P_2 - P_3)t^2 \\
& + (-P_0 + 3P_1 - 3P_2 + P_3)t^3 \Big]
\end{aligned}
\end{equation}

\noindent
where 

\begin{itemize}
  \item \( C(t) \) is the position on the time. \( t \).
  \item \( P_0, P_1, P_2, P_3 \) are control points.
  \item \( t \) ranges from 0 to 1, interpolating between \( P_1 \) and \( P_2 \).
\end{itemize}

This equation ensures that at \( t = 0 \), \( C(t) = P_1 \), the curve starts at the second control point, whereas at \( t = 1 \), \( C(t) = P_2 \), the curve ends at the third control point.

\begin{algorithm*}[t]
\footnotesize
\caption{extract\_road\_geometry(\textit{opendrive, lane\_side})}
\label{algorithm1}
\begin{algorithmic}[1]
\State \textit{road\_data} $\gets$ [] \Comment{Initialize a list to store road data}

\For{each \textit{road} in \textit{opendrive}} 
    \State \textit{planView} $\gets$ \textit{road}.planView
    \For{each \textit{geometry} in \textit{planView}.geometries} \Comment{Iterate through geometries of the road}
        \State $(x, y, s)$ $\gets$ $parse\_geometry$(geometry) \Comment{Extract geometry attributes}
        \State \textit{elevation\_width} $\gets$ \textit{get\_elevation\_and\_width}(road, s)

            \State \textit{lane\_offset} $\gets$ \textit{compute\_lane\_offset}({road}, $s$,{lane\_side}) \Comment{lane\_side is either left or right}

            \State \textit{adjusted\_x} $\gets$ $x$ + \textit{lane\_offset}.x \Comment{Adjust x-coordinate for lane}
            \State \textit{adjusted\_y} $\gets$ $y$ + \textit{lane\_offset}.y \Comment{Adjust y-coordinate for lane}

            \State \textit{append} (adjusted\_x, adjusted\_y, {elevation\_width.z}, {elevation\_width.width}) to \textit{road\_data}  
    \EndFor
\EndFor
\State \textbf{return} \textit{road\_data}
\end{algorithmic}
\end{algorithm*}

To create a complete spline, this process is repeated for each pair of points, connecting small segments smoothly. The advantages of this spline include its local control, smoothness, and computational efficiency. However, if the control points are unevenly spaced, the curve can overshoot. Also, while the curve is smooth, it does not always guarantee a perfect curvature at each point, which can be a limitation in precision-demanding applications like robotics. Despite these minor drawbacks, Catmull-Rom splines remain adequate for generating road topologies in a versatile manner~\cite{catmull-rom}.

\begin{figure}[!tbp]
\centering
\includegraphics[width=0.49\textwidth]{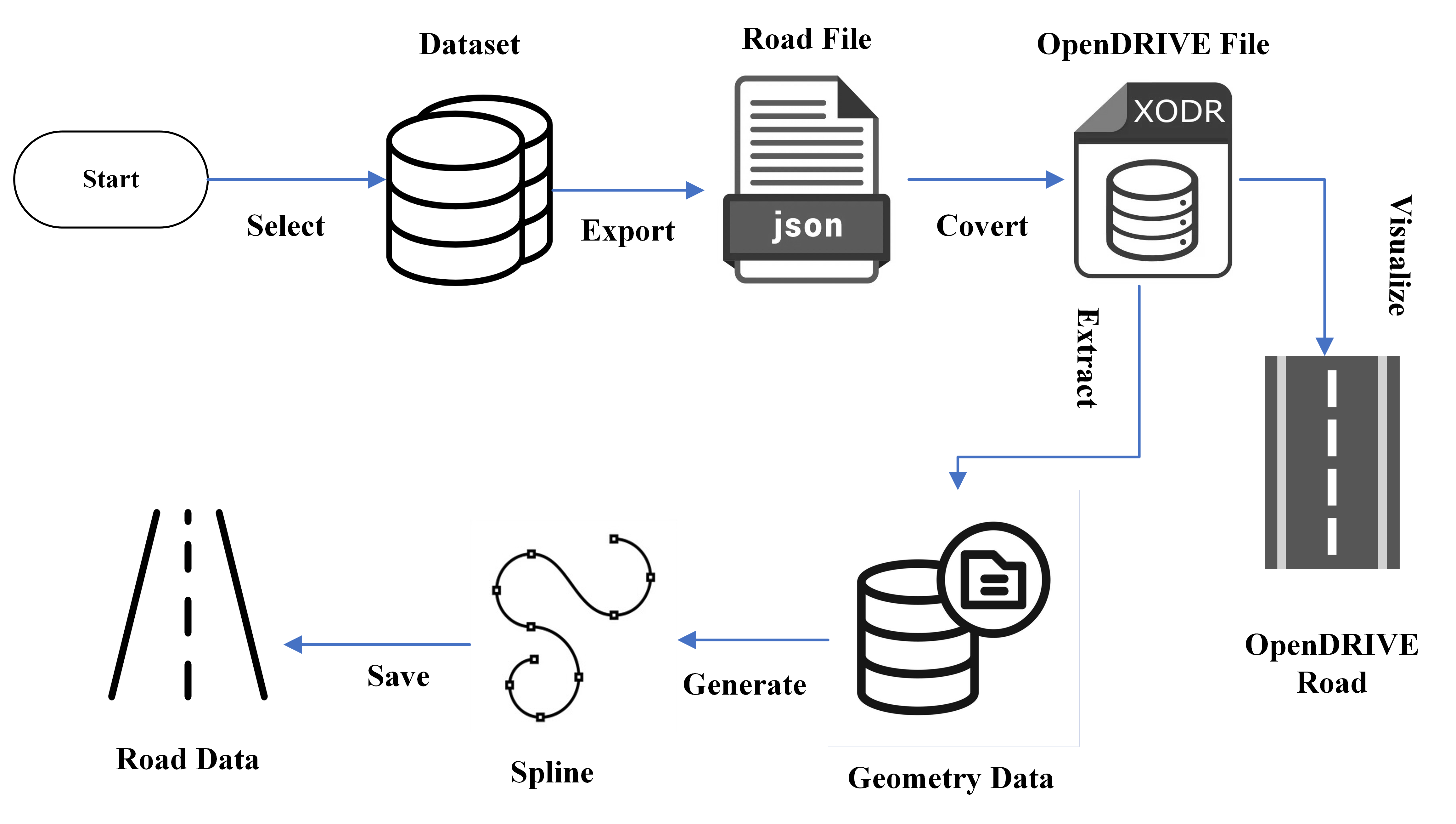}
% \vspace{-8mm}
\caption{\label{fig:workflow} Conversion process from OpenDRIVE format to Catmull-Rom spline.}
\end{figure}

\section{OpenCat: from OpenDRIVE to Catmull Rom spline} \label{sec:conversion}

This section provides an overview of the road conversion process from OpenDRIVE to Catmull-Rom spline. The conversion process involves simplifying the detailed road model to focus on its geometry, specifically its shape and path. The diagram in Figure~\ref{fig:workflow} illustrates the conversion process. First, we export the OpenDRIVE road field from the dataset scenario files in JSON format. These JSON files were then converted to OpenDRIVE (.xodr) files used by \textsc{OpenCat}. 
After that, we extract the necessary geometry attributes. Then, we calculate the reference line (centerline) from the OpenDRIVE data, which involves averaging lane boundaries (right and left). The OpenDRIVE road network is built around a central reference line, which acts as the core element of every road. The reference line defines how the road bends and turns from the planView. All lanes and elevation profiles are linked to this reference line as shown in Figure~\ref{fig:rline}. 

Next, the specific points that define a spline known as control points are defined along this reference line to generate the Catmull-Rom spline, ensuring that they are spaced to accurately capture the road's curvature. 
Additionally, we integrate elevation data from the OpenDRIVE file by applying the elevation profile to the spline, although this step is challenging because splines typically do not directly account for such details. The whole process of data extraction and spline generation is explained next.

\begin{algorithm*}[t]
\footnotesize
\caption{generate\_spline(\textit{opendrive, lane\_side})}
\label{algorithm2}
\begin{algorithmic}[1]

\State \textit{right\_lane\_points} $\gets$ extract\_road\_geometry(\textit{opendrive, right})  \Comment{Extract right lane points}
\State \textit{left\_lane\_points} $\gets$ extract\_road\_geometry(\textit{opendrive, left})  \Comment{Extract left lane points}
\State \textit{min\_points} $\gets 4$ \Comment{Minimum points for spline generation}
\If{len (\textit{right\_lane\_points})  \textbf{or} len (\textit{left\_lane\_points}) $< \textit{min\_points}$}
    \State \textbf{return} $empty\_list$ 
\EndIf

\State \textit{control\_points} $\gets$ \textit{compute\_centerline}({right\_lane\_points, left\_lane\_points}) 
\Comment{Compute centerline}
\State \textit{spline\_generator} $\gets$ \textit{compute_catmull_rom_spline}({control\_points}) 
\Comment{Generate spline}
   
\State \textit{spline\_points} $\gets$ \textit{spline\_generator}.generate\_spline() 

\State \textbf{return} \textit{{control\_points}, {spline\_points}}
   
\end{algorithmic}
\end{algorithm*}

\subsection{Conversion Algorithm} \label{subsec:implementation} 

\textit{Algorithm~\ref{algorithm1}} is designed to process an OpenDRIVE file and extract the road geometry for a specified lane side (right, left), which will be used to generate a spline. For each road in an OpenDrive file, the algorithm iterates through its geometry sections to extract the attributes such as position (\textit{x}, \textit{y}), and starting position (\textit{s}) by calling the auxiliary function \textit{parse\_geometry} (Line~5). Next, it calculates the elevation profile and width of the road at the current position \textit{`s'} based on polynomial coefficients found in the OpenDRIVE data (Line~6). Then we compute the lane offset (Line~7), which is the total width of the lanes from the road center, depending on the specified lane side (left or right). Based on the \textit{lane_side} parameter, the algorithm adjusts the \textit{x} and \textit{y} coordinates by the computed lane offset (Lines~8-9). This adjustment ensures that only the points from the desired lane side are extracted. 

Finally, the extracted road data \textit{(x, y, z, width)} are collected (line 10) and returned (line 13). Once the road geometry has been fully extracted, we proceed to generate the spline curve using \textit{Algorithm~\ref{algorithm2}}, which constructs a Catmull-Rom spline and interpolates points along the spline, ensuring that the road geometry is represented as a smooth curve. Details of the supporting functions used in this process are explained below.

\vspace{0.3\baselineskip}
\begin{enumerate} [leftmargin=5pt, labelwidth=0pt, labelsep=0.5em, align=left]
    \item \textit{Lateral and Elevation Profile}
\end{enumerate}
\vspace{0.3\baselineskip}

The \textit{parse\_geometry} function is designed to extract the basic geometry attributes of a road from the OpenDRIVE XML structure. It parses the geometry section of a road, which contains essential attributes such as \textit{`x'} and \textit{`y'} coordinates, \textit{`length'} (the length of the geometry section), \textit{`hdg'} (the heading or direction of the road at the starting point), and \textit{`s'} (the position along the road where the geometry section begins). These values provide the fundamental data that describes the shape and orientation of the road section. This information is used to calculate the road geometry and lane offsets.

Although the data extracted through the parse\_geometry function is sufficient for generating a flat 2D plane that represents the road's horizontal geometry, it does not capture the true 3D nature of the road terrain, such as slopes. To address this, we use the \textit{get\_elevation\_and\_width} function to extract both the elevation and the width of the road at a specified position \textit{`s'}, ensuring a more accurate road representation. If the current road has an associated elevation profile, it iterates over the elevation segments and calculates the elevation for the given position (s) using polynomial coefficients. For each segment, it checks whether the current position \textit{`s'} lies within that segment by comparing \textit{`s'} with the starting point \textit{`s\_elev'} of the elevation and the length of the segment. If \textit{`s'} falls within this range, the function calculates the elevation \textit{`z'} at this position using a cubic polynomial formula:

\begin{figure*}[!tbp]
    \centering
    % Subfigure a
    \subfigure[\parbox{.9\linewidth}{Original OpenDRIVE road.}]{
        % \fbox{ % trim={<left> <lower> <right> <upper>}
        \includegraphics[width=0.3\textwidth, height=4cm]{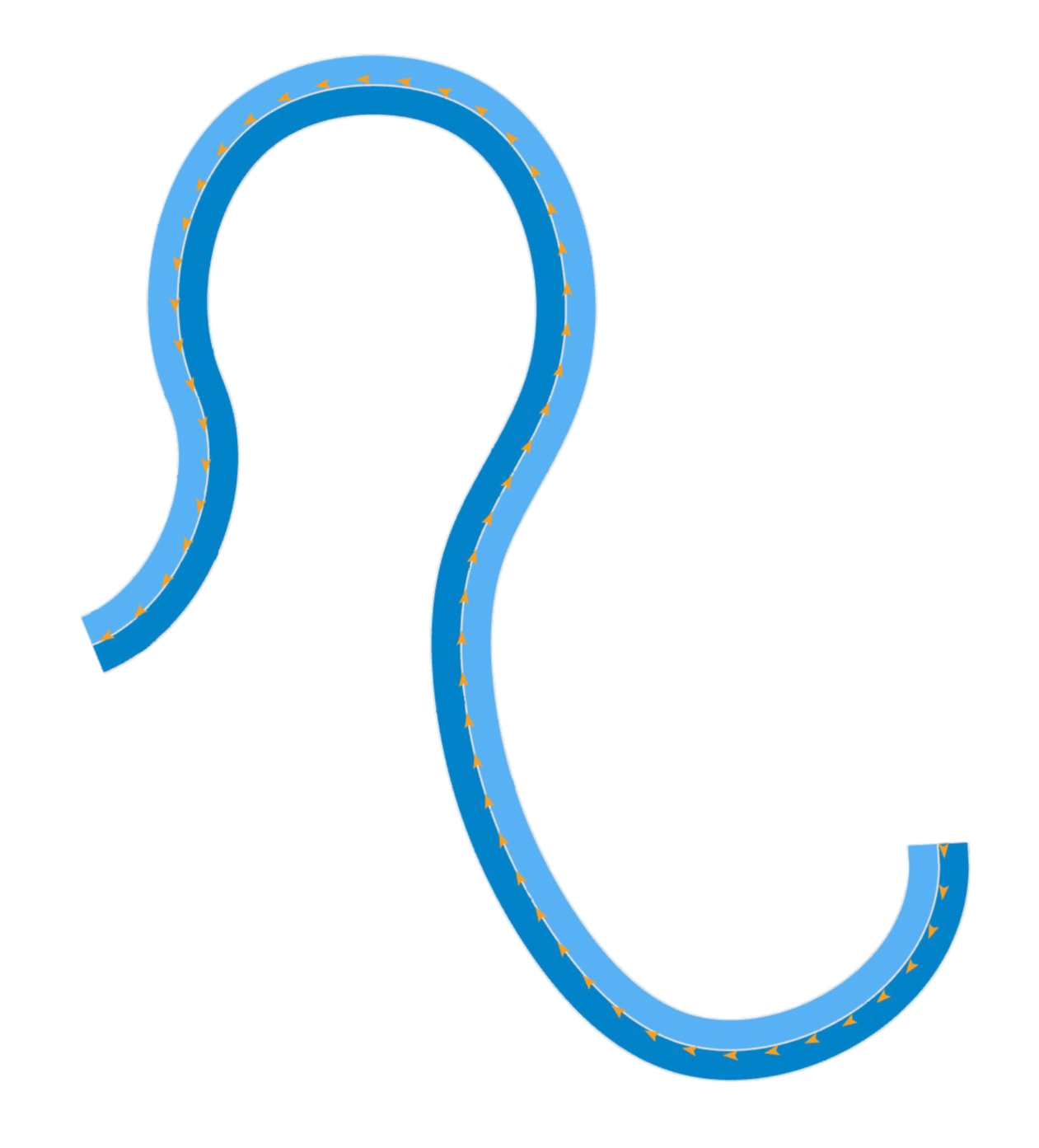}
        % }
        \label{fig:oproad}
    }
    % Subfigure b
    \subfigure[\parbox{.8\linewidth}{Generated spline with original points.}]{
        \includegraphics[width=0.3\textwidth, height=5cm, keepaspectratio]{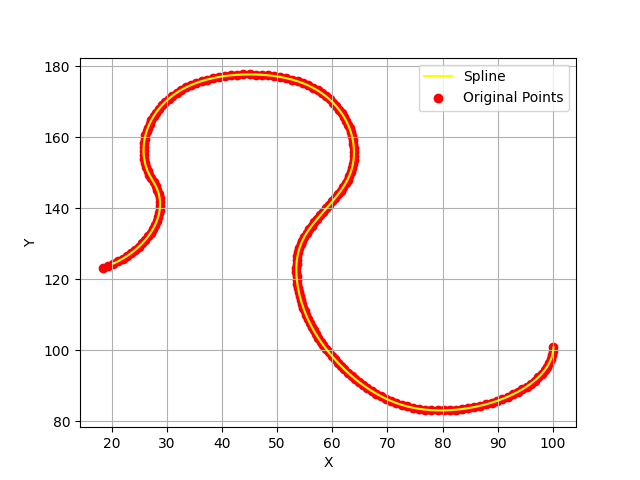}
        \label{fig:spline}
    }
    % Subfigure c
    \subfigure[\parbox{.9\linewidth}{Road in Udacity simulator.}]{
        \includegraphics[width=0.3\textwidth, height=4.5cm, keepaspectratio]{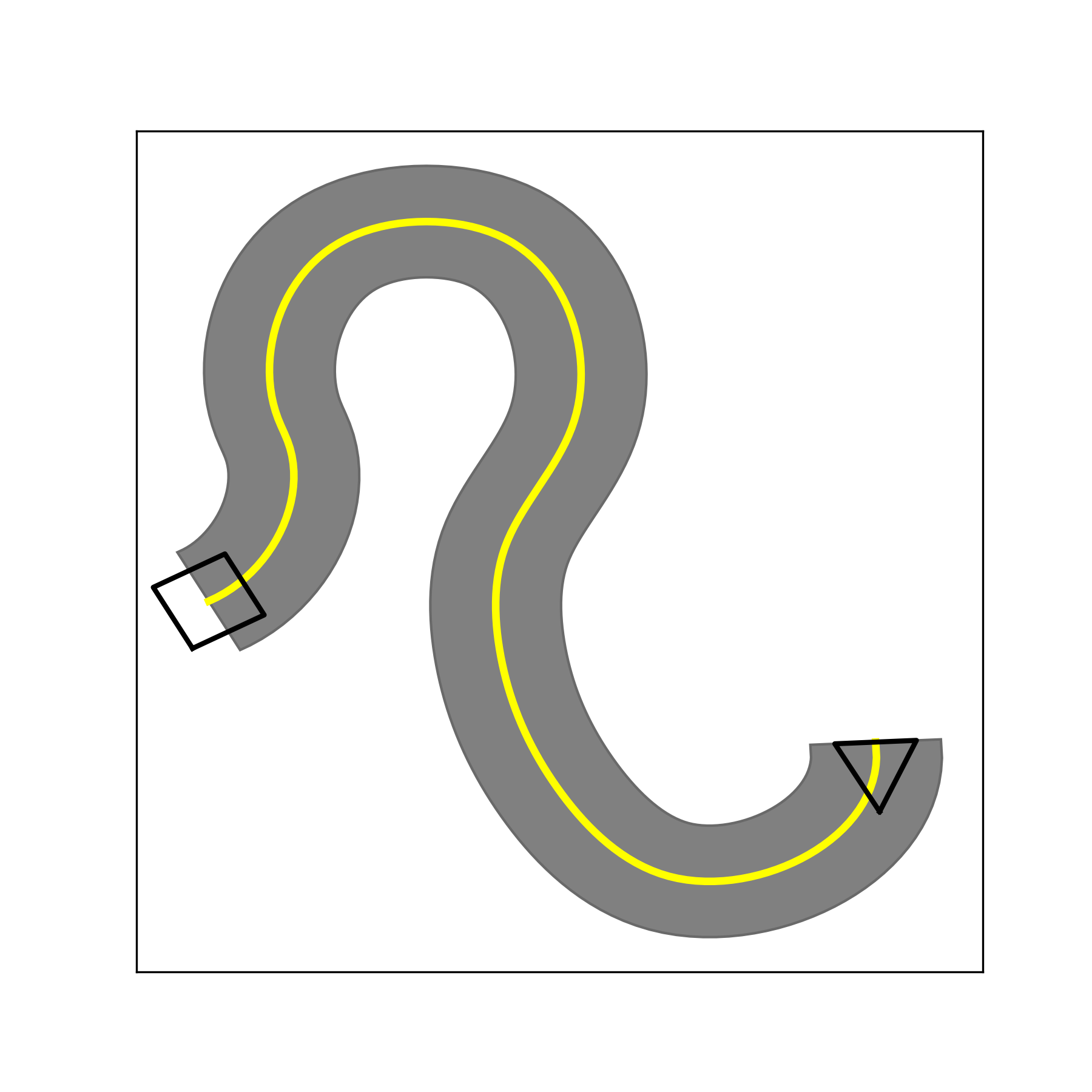}
        \label{fig:test}
    }
    \caption{Comparative view of OpenDRIVE road, generated spline, and test case.}
    % \vspace{-4mm}
\end{figure*}

\begin{equation}
\label{eq2}
z = a + b \cdot ds + c \cdot ds^2 + d \cdot ds^3
\end{equation}

Here \textit{(a, b, c, d)} are polynomial coefficients, and \textit{`ds'} is the difference between \textit{`s'} and the start of the segment \textit{(ds = s - s\_elev)}.
After calculating the elevation, the function extracts the road's width by locating the lanes within the \textit{`laneSection'}. For each lane in the section, it calculates the width using similar polynomial coefficients and sums the widths to get the total road width. The calculated elevation and width are then returned, providing the necessary data to represent the physical characteristics of the road at that position.

\vspace{0.3\baselineskip}
\begin{enumerate}[resume, leftmargin=5pt, labelwidth=0pt, labelsep=0.5em, align=left]
    \item \textit{Lane Offset}
\end{enumerate}
\vspace{0.3\baselineskip}

The \textit{compute\_lane\_offset} function determines the lateral positioning (side-to-side) of the lanes on the road at a given point (s) along the road geometry. It is essential for accurate road modeling, simulation, and visualization, especially when dealing with road curvature or varying lane widths. 

First, we check the validity of each lane in the lane section on the road, skipping over the center lane \textit{(lane\_id = 0)}. For valid lanes, the width of each lane is calculated using polynomial coefficients, much like the elevation and width calculations earlier. The width of each lane is then added to the \textit{‘lane\_offset’}, which accumulates the total offset of all valid lanes for the specified side. Once all lanes have been processed, the total lane offset is returned.

Given the previous information, we now have details on how to generate a spline. 
\textit{Algorithm~\ref{algorithm2}} generates a smooth spline (curve) representing the road's reference line (centerline). It begins by calling \textit{Algorithm~\ref{algorithm1}} to extract road geometry data for the specified lane side (\textit{right} or \textit{left}) from the OpenDRIVE file (Lines~1-2). These points represent the lane positions along the road. Next, the algorithm ensures there are enough points to generate a spline by checking if both the right and left lane data meet the minimum point requirement (Lines~3). If there are enough points, the algorithm proceeds by calculating the road's centerline by averaging the points of the right and left lanes to determine a set of control points that represent the middle of the road (Line~7). 
With these control points in hand, the algorithm initializes a \textit{`CatmullRomSpline'} generator, which creates a smooth curve (spline) based on the control points (Line~8). To control the smoothness of the curve, the generator is configured with an \textit{`alpha'} value of 0.5 and is set to produce one spline point per segment. Finally, the algorithm generates the spline points and returns both the control points and the generated spline points (Lines~9-10). 

\section{Experiments and Results} \label{sec:results}

To evaluate the performance of \textsc{OpenCat}, we applied it to the publicly available OpenDRIVE scenario dataset called \textsc{SensoDat}~\cite{sensodat}. More in detail, we aim to answer the following research questions:

\noindent \textbf{RQ\textsubscript{1} (Accuracy)}: How accurately does \textsc{OpenCat} transform the OpenDRIVE road geometry into the Catmull-Rom spline representation?

\noindent \textbf{RQ\textsubscript{2} (Comparison)}: How does the converted dataset perform relative to the original \textsc{SensoDat} dataset in terms of the test pass/fail ratio?

\subsection{RQ\textsubscript{1} (Accuracy)} 

We converted a total of 32,580 roads from the dataset, but for comparison, we visually present only a single example of conversion; other results can be verified at the publicly available GitHub repository~\cite{opendrive2catmullrom}. Before converting the OpenDRIVE files into the spline, we verified the validity and usability of the data in the OpenDRIVE files by visualizing the road structure using the OpenDRIVE file viewers~\cite{libOpenDRIVE,pagel_odrviewer}. Figure~\ref{fig:oproad} presents an original road from the OpenDRIVE scenario visualized through the OpenDRIVE viewer~\cite{libOpenDRIVE}, while Figure~\ref{fig:spline} presents the conversion result, where the generated spline (yellow) is overlaid with the original points from the OpenDRIVE file (red dots). We can see that the generated spline closely follows the original geometry, demonstrating the accuracy of the conversion. The visual results confirm that \textsc{OpenCat} preserves the original geometric characteristics of the road while producing a smooth representation suitable for use in simulations.

%\subsection{Accuracy Metrics} %\label{subsec:accuracy}
In addition to visual validation, we also verify our results using the following metrics to measure the precision of the generated spline by comparing the spline points with the original points.
The accuracy metric evaluates how closely the spline interpolation matches the original points by calculating the average Euclidean distance between the spline points and the original points. We then normalize this average distance by the maximum possible error, defined as the diagonal distance of the bounding box that encloses the original points. The resulting percentage indicates how accurate the spline interpolation is with respect to the original points:

\begin{equation}
\label{eq3}
\text{Accuracy} = \left(1 - \frac{\text{Average Distance}}{\text{Max Possible Error}}\right) \times 100
\end{equation}

%\subsubsection{R-square value} 
R-squared (coefficient of determination) quantifies the goodness of fit of the spline interpolation compared to the original data. It measures how well the spline points fit the original data. An R² value closer to 1 indicates that the spline fits the data well, while lower values signify a poor fit.

\begin{equation}
\label{eq4}
R^2 = 1 - \frac{\sum (y_i - \hat{y}_i)^2}{\sum (y_i - \bar{y})^2}
\end{equation}

where \( y_i \) are the actual values. \( \hat{y}_i \) are the predicted/interpolated values and \( \bar{y} \) is the mean of the original values.
The conversion achieved perfect accuracy (accuracy $= 100\%$) and R-squared ($R^2$=1), confirming the correctness and quality of \textsc{OpenCat}.

\subsection {RQ\textsubscript{2} (Comparison)}

We implemented and validated our spline-based roads for the Udacity simulator~\cite{udacity_sim} using the test case generation framework by Biagiola et al.~\cite{biagiola2024}. The framework uses a DAVE-2 model as a system under test~\cite{nvidia-dave2}, which is an advanced driver assistance system for lane-keeping. 

First, we initialize road scenarios by defining control points derived from our generated spline, similar to the test case representation in Biagiola et al.~\cite{biagiola2024}. The control points represent critical aspects of road geometry from the OpenDRIVE file, such as lane boundaries, curves, and elevation changes. After interpolation, a road is considered valid if it meets certain criteria, which include: (1) the start and end points are distinct; (2) the road remains within a predefined square bounding box of 250 × 250 units; and (3) there are no intersections. Figure~\ref{fig:test} presents an example of such a valid road generated by the Udacity test generator using our road data. 
We executed simulations on a MacBook Pro M2, 8 cores, 16 GB RAM, macOS Sequoia, and on an Intel Core i5, 12 cores, 16 threads, 4GB DDR6 RAM, NVIDIA GeForce RTX 3050 Ti, 16 GB, Windows 11. 

\begin{table}[t]
\scriptsize
\centering
\caption{Test pass/fail counts across campaigns.}
\resizebox{\columnwidth}{!}{%
\begin{tabular}{*{7}{c}}
\toprule

\textbf{Campaign} & \textbf{Pass} & \textbf{Fail} & \textbf{Total} & \textbf{Exe\_Time} & \textbf{\textsc{OpenCat}} & \textbf{\textsc{SensoDat}}\\
 & & & & \textbf{h/m} & \textbf{Pass\%} & \textbf{Pass\%} \\

\midrule
\multicolumn{7}{c}{\textbf{Ambeigen Campaigns}} \\
\rowcolor{green!20} 2\_ambiegen(w) & 963 & 10 & 973 & 5/19 & 98.97 & 58.17 \\ % High difference (Good)
\rowcolor{green!20} 3\_ambiegen(w) & 964 & 0 & 964 & 5/41 & 100 & 53.32 \\
\rowcolor{green!20} 4\_ambiegen(m) & 956 & 9 & 965 & 5/50 & 99.06 & 51.30 \\
\rowcolor{green!20} 5\_ambiegen(m) & 952 & 6 & 958 & 5/23 & 99.37 & 56.68 \\
\rowcolor{green!20} 6\_ambiegen(m) & 951 & 8 & 959 & 5/44 & 99.16 & 54.12 \\
\rowcolor{green!20} 7\_ambiegen(m) & 952 & 11 & 963 & 5/46 & 98.85 & 53.80 \\
\rowcolor{green!20} 8\_ambiegen(m) & 941 & 11 & 952 & 5/46 & 98.84 & 54.31 \\
\rowcolor{green!20} 9\_ambiegen(m) & 944 & 9 & 953 & 5/57 & 99.05 & 50.79 \\
\rowcolor{green!20} 10\_ambiegen(m) & 964 & 7 & 971 & 5/38 & 99.27 & 53.88 \\
\rowcolor{green!20} 11\_ambiegen(m) & 966 & 7 & 973 & 5/37 & 99.28 & 50.57 \\
\rowcolor{green!20} 13\_ambiegen(w) & 944 & 10 & 954 & 5/34 & 98.95 & 53.85 \\
\rowcolor{green!20} 14\_ambiegen(m) & 951 & 8 & 959 & 5/43 & 99.16 & 56.94 \\
\rowcolor{green!20} 15\_ambiegen(m) & 939 & 13 & 952 & 5/33 & 98.63 & 54.20 \\

\addlinespace
\multicolumn{7}{c}{\textbf{Frenetic Campaigns}} \\
\rowcolor{green!20} 2\_frenetic(m) & 920 & 8 & 928 & 5/38 & 99.13 & 60.56 \\
\rowcolor{green!20} 3\_frenetic(m) & 944 & 10 & 954 & 5/36 & 98.95 & 63.34 \\
\rowcolor{green!20} 4\_frenetic(m) & 956 & 8 & 964 & 5/44 & 99.17 & 65.06 \\
\rowcolor{green!20} 5\_frenetic(m) & 935 & 10 & 945 & 5/39 & 98.94 & 66.67 \\
\rowcolor{red!20} 6\_frenetic(w) & 854 & 90 & 944 & 5/33 & 90.46 & 65.93 \\  % Low difference (Bad)
\rowcolor{green!20}7\_frenetic(m) & 958 & 9 & 967 & 5/53 & 99.06 & 64.33 \\
\rowcolor{yellow!20} 8\_frenetic(w) & 859 & 93 & 952 & 5/37 & 94.01 & 63.87 \\  % Moderate difference
\rowcolor{green!20}9\_frenetic(m) & 959 & 5 & 964 & 5/51 & 99.48 & 64.20 \\
\rowcolor{green!20} 11\_frenetic(w) & 828 & 28 & 866 & 5/17 & 96.76 & 63.51 \\
\rowcolor{green!20} 12\_frenetic(w) & 928 & 28 & 956 & 5/38 & 97.07 & 63.07 \\
\rowcolor{yellow!20} 13\_frenetic(w) & 925 & 34 & 959 & 5/38 & 96.45 & 67.06 \\
\rowcolor{yellow!20} 14\_frenetic(m) & 857 & 9 & 866 & 5/10 & 98.96 & 68.59 \\
\rowcolor{green!20} 15\_frenetic(m) & 860 & 10 & 870 & 5/5 & 98.87 & 61.84 \\

\addlinespace
\multicolumn{7}{c}{\textbf{Frenetic\_v Campaigns}} \\

\rowcolor{green!20} 2\_frenetic\_v(w) & 943 & 1 & 944 & 5/51 & 99.90 & 67.63 \\
\rowcolor{green!20} 4\_frenetic\_v(m) & 524 & 1 & 525 & 3/16 & 99.80 & 65.52 \\
\rowcolor{green!20} 5\_frenetic\_v(w) & 940 & 0 & 940 & 5/49 & 100 & 67.45 \\
\rowcolor{green!20} 6\_frenetic\_v(w) & 763 & 1 & 764 & 4/42 & 99.86 & 66.58 \\
\rowcolor{green!20} 7\_frenetic\_v(w) & 47 & 0 & 47 & 0/18 & 100 & 55.32 \\
\rowcolor{green!20} 11\_frenetic\_v(w) & 949 & 4 & 953 & 5/58 & 99.58 & 65.27 \\
\rowcolor{yellow!20} 12\_frenetic\_v(w) & 888 & 54 & 942 & 5/53 & 94.26 & 66.24 \\
\rowcolor{yellow!20}13\_frenetic\_v(m) & 946 & 5 & 951 & 6/1 & 99.47 & 70.97 \\
\rowcolor{yellow!20}14\_frenetic\_v(m) & 928 & 6 & 934 & 5/44 & 99.35 & 72.93 \\
\rowcolor{green!20} 15\_frenetic\_v(w) & 925 & 24 & 949 & 5/58 & 97.47 & 64.18 \\

\bottomrule

\end{tabular}
}
\label{tab:results}
\end{table}

Table~\ref{tab:results} specifies the campaigns executed on either macOS or Windows, with (m) indicating macOS and (w) indicating Windows. This dual-system approach also helps further validate our methodology by testing the \textsc{OpenCat}'s converted dataset performance on two different operating systems.

A test case is considered successful if the vehicle stays within the correct lane (right) up to the final road control point. In contrast, a test case fails if the car goes out of bounds (OOB), i.e., off the road
\cite{gambi2019automatically}, or if the road's endpoint overlaps with the starting point (i.e., road\_818 from campaign\_2\_ambiegen), which prevents the simulator from starting, consequently failing the test case. 

Each scenario began with the vehicle positioned at the starting control point, centered within the lane, and orientated to match the direction of the road, which helped establish a consistent starting condition across tests. 
As the vehicle drives along the paths defined by the spline, we monitored essential metrics to ensure realistic driving behavior, as outlined in previous work~\cite{biagiola2024}. These metrics included lane-keeping accuracy, which checked the vehicle's lateral position to ensure it remained within lane boundaries, and OOB detection. This process allowed us to track safety violations~\cite{biagiola2024}. 

Table~\ref{tab:results} presents the simulation results from our simulation; we executed 32,580 tests in total. The results of our experiments reveal a significant improvement in the ratio of passing tests across all test campaigns compared to the original \textsc{SensoDat}~\cite{sensodat} based on the out\_of\_bound (OOB) metric. 
The table employs a color scheme to distinguish the comparison between roads generated through \textsc{OpenCat}'s translation and those provided by \textsc{SensoDat}. The green color indicates campaigns where \textsc{OpenCat} outperformed \textsc{SensoDat} with a percentage difference greater than 30\%, the yellow color represents campaigns with a difference of 25\% to 30\%, and the red color indicates cases where the difference was less than 25\%. 
Unlike the original dataset, where pass rates varied considerably—especially in challenging road structures such as Frenetic campaigns—our results show consistently high pass rates, with 32,035 passing tests and 545 failing tests, resulting in an overall pass ratio of 98\%, compared to a 61\% pass ratio of the original \textsc{SensoDat} data set, which recorded 19,926 passing tests and 12,654 failing tests.

In particular, while the Frenetic campaigns in \textsc{SensoDat} displayed several failure rates, our results show that an independent ADAS model achieves substantial success rates, even the most challenging campaigns achieving over 95\% pass rates, except for a few specific cases, such as campaign\_6\_frenetic and campaign\_8\_frenetic. Ambiegen campaigns~\cite{HUMENIUK2023102990}, which performed very low in the original dataset, performed exceptionally well in our results, with pass rates consistently exceeding 98\%, compared to the range of 50\% to 58\%, while macOS and Windows executions delivered similar results, highlighting the robustness of our evaluation. Similarly, FreneticV campaigns, which generally showed strong pass rates in \textsc{SensoDat}, displayed near-ideal results in our simulations. By including execution time as a metric, we provide insight into the computational efficiency of different campaigns, offering valuable information to optimize resource allocation in future simulations. 

Our findings demonstrate significant improvements in test pass rates compared to \textsc{SensoDat} and reveal an important limitation in current ADAS benchmarks. While the converted geometry is identical to the original, as established in RQ\textsubscript{1}, the significant difference in test pass/fail results can be attributed to the interaction between the representation format, the ADAS under test (e.g., Dave-2), and the simulator environment. 

Future work will investigate the specific factors contributing to the observed results. Moreover, benchmarks like \textsc{SensoDat} are coupled to the specific ADAS model with which they were created, limiting their utility in evaluating alternative ADAS systems. The ADAS under test, independent of \textsc{SensoDat}, successfully navigates most roads with minimal issues, highlighting the need for architecture-agnostic benchmarks for regression testing purposes. We propose the development of ADS-independent benchmarks that enable meaningful comparisons across diverse ADAS models and systems. This represents an open research problem in ADAS testing and a critical area for future investigation.

\section{Conclusions} \label{sec:conclusion}

In this study, we introduced \textsc{OpenCat}, a novel tool designed to convert the OpenDRIVE road format into Catmull-Rom spline representations, helping to bridge the gap in road format interoperability for autonomous driving applications. 

Our converter \textsc{OpenCat} simplifies the complexities of generating precise road maps, making them adaptable for simulation to test lane-keeping ADAS models. By applying it to the \textsc{SensoDat} dataset, we demonstrated its effectiveness in accurately translating OpenDRIVE road geometry into realistic and smooth road scenarios. The integration of converted roads into the Udacity self-driving car simulator further validated \textsc{OpenCat}'s capacity to generate reliable, simulation-ready roads suitable for diverse autonomous driving test environments. Our study demonstrates that ADAS under test, independent of \textsc{SensoDat}, successfully navigates a significant portion of the roads (98\%) without any failure. This indicates that the effectiveness of \textsc{SensoDat} as a regression testing benchmark may be limited and we propose developing architecture-agnostic benchmarks as a crucial step toward advancing ADAS testing and evaluation.

\section*{Acknowledgements}
\addcontentsline{toc}{section}{Acknowledgements}

This work has been partially supported by the Centro Nazionale HPC, Big Data e Quantum Computing (PNRR CN1 spoke 9 Digital Society \& Smart Cities); the Engineered MachinE Learning-intensive IoT systems (EMELIOT) national research project, which has been funded by the MUR under the PRIN 2020 program (Contract 2020W3A5FY); the MUR under the grant ``Dipartimenti di Eccellenza 2023-2027'' of the Department of Informatics, Systems, Communication of the University of Milano-Bicocca, Italy, and the Bavarian Ministry of Economic Affairs, Regional Development and Energy.

\balance 
\bibliographystyle{IEEEtran}
\bibliography{references}

\begin{thebibliography}{10}
\providecommand{\url}[1]{#1}
\csname url@samestyle\endcsname
\providecommand{\newblock}{\relax}
\providecommand{\bibinfo}[2]{#2}
\providecommand{\BIBentrySTDinterwordspacing}{\spaceskip=0pt\relax}
\providecommand{\BIBentryALTinterwordstretchfactor}{4}
\providecommand{\BIBentryALTinterwordspacing}{\spaceskip=\fontdimen2\font plus
\BIBentryALTinterwordstretchfactor\fontdimen3\font minus \fontdimen4\font\relax}
\providecommand{\BIBforeignlanguage}[2]{{%
\expandafter\ifx\csname l@#1\endcsname\relax
\typeout{** WARNING: IEEEtran.bst: No hyphenation pattern has been}%
\typeout{** loaded for the language `#1'. Using the pattern for}%
\typeout{** the default language instead.}%
\else
\language=\csname l@#1\endcsname
\fi
#2}}
\providecommand{\BIBdecl}{\relax}
\BIBdecl

\bibitem{survey-lei-ma}
S.~Tang, Z.~Zhang, Y.~Zhang, J.~Zhou, Y.~Guo, S.~Liu, S.~Guo, Y.-F. Li, L.~Ma, Y.~Xue, and Y.~Liu, ``{A Survey on Automated Driving System Testing: Landscapes and Trends},'' \emph{ACM Trans. Softw. Eng. Methodol.}, vol.~32, no.~5, Jul. 2023.

\bibitem{yurtsever2020survey}
E.~Yurtsever, J.~Lambert, A.~Carballo, and K.~Takeda, ``A survey of autonomous driving: Common practices and emerging technologies,'' \emph{IEEE access}, vol.~8, pp. 58\,443--58\,469, 2020.

\bibitem{li2024panopticperceptionautonomousdriving}
\BIBentryALTinterwordspacing
Y.~Li and L.~Xu, ``Panoptic perception for autonomous driving: A survey,'' 2024. [Online]. Available: \url{https://arxiv.org/abs/2408.15388}
\BIBentrySTDinterwordspacing

\bibitem{grigorescu2020survey}
S.~Grigorescu, B.~Trasnea, T.~Cocias, and G.~Macesanu, ``A survey of deep learning techniques for autonomous driving,'' \emph{Journal of Field Robotics}, vol.~37, no.~3, pp. 362--386, 2020.

\bibitem{2020-Riccio-EMSE}
V.~Riccio, G.~Jahangirova, A.~Stocco, N.~Humbatova, M.~Weiss, and P.~Tonella, ``{Testing Machine Learning based Systems: A Systematic Mapping},'' \emph{Empirical Software Engineering}, 2020.

\bibitem{Elghazaly2023HighDefinitionMC}
\BIBentryALTinterwordspacing
G.~Elghazaly, R.~Frank, S.~Harvey, and S.~Safko, ``High-definition maps: Comprehensive survey, challenges, and future perspectives,'' \emph{IEEE Open Journal of Intelligent Transportation Systems}, vol.~4, pp. 527--550, 2023. [Online]. Available: \url{https://api.semanticscholar.org/CorpusID:259915452}
\BIBentrySTDinterwordspacing

\bibitem{banafa2024}
A.~Banafa, ``Technical challenges in autonomous vehicle development,'' \url{https://www.linkedin.com/pulse/technical-challenges-autonomous-vehicle-development-banafa-ypfhc/}, 2024, accessed: 2024-11-03.

\bibitem{2022-Stocco-ASE}
A.~Stocco, P.~J. Nunes, M.~d'Amorim, and P.~Tonella, ``{ThirdEye: Attention Maps for Safe Autonomous Driving Systems},'' in \emph{Proceedings of 37th IEEE/ACM International Conference on Automated Software Engineering}, ser. ASE '22.\hskip 1em plus 0.5em minus 0.4em\relax IEEE/ACM, 2022.

\bibitem{2020-Stocco-ICSE}
A.~Stocco, M.~Weiss, M.~Calzana, and P.~Tonella, ``Misbehaviour prediction for autonomous driving systems,'' in \emph{Proceedings of ACM 42nd International Conference on Software Engineering}, ser. ICSE '20.\hskip 1em plus 0.5em minus 0.4em\relax ACM, 2020, p. 12 pages.

\bibitem{2023-Stocco-EMSE}
A.~Stocco, B.~Pulfer, and P.~Tonella, ``Model vs system level testing of autonomous driving systems: a replication and extension study,'' \emph{Empirical Software Engineering}, vol.~28, no.~3, p.~73, May 2023.

\bibitem{2021-Stocco-JSEP}
\BIBentryALTinterwordspacing
A.~Stocco and P.~Tonella, ``Confidence-driven weighted retraining for predicting safety-critical failures in autonomous driving systems,'' \emph{Journal of Software: Evolution and Process}, 2021. [Online]. Available: \url{https://doi.org/10.1002/smr.2386}
\BIBentrySTDinterwordspacing

\bibitem{2024-Lambertenghi-ICST}
S.~C. Lambertenghi and A.~Stocco, ``Assessing quality metrics for neural reality gap input mitigation in autonomous driving testing,'' in \emph{Proceedings of the IEEE 17th International Conference on Software Testing, Verification and Validation}, ser. ICST '24.\hskip 1em plus 0.5em minus 0.4em\relax IEEE, 2024, p. 12 pages.

\bibitem{2025-Lambertenghi-ICST}
S.~C. Lambertenghi, H.~Leonhard, and A.~Stocco, ``Benchmarking image perturbations for testing automated driving assistance systems,'' in \emph{Proceedings of the IEEE 18th IEEE International Conference on Software Testing, Verification and Validation}, ser. ICST '25.\hskip 1em plus 0.5em minus 0.4em\relax IEEE, 2025, p. 12 pages.

\bibitem{2023-Stocco-TSE}
A.~Stocco, B.~Pulfer, and P.~Tonella, ``{Mind the Gap! A Study on the Transferability of Virtual Versus Physical-World Testing of Autonomous Driving Systems},'' \emph{IEEE Transactions on Software Engineering}, vol.~49, no.~04, pp. 1928--1940, apr 2023.

\bibitem{kim2017testing}
B.~Kim, Y.~Kashiba, S.~Dai, and S.~Shiraishi, ``Testing autonomous vehicle software in the virtual prototyping environment,'' \emph{IEEE Embedded Systems Letters}, vol.~9, no.~1, pp. 5--8, 2017.

\bibitem{zofka2016testing}
M.~R. Zofka, S.~Klemm, F.~Kuhnt, T.~Schamm, and J.~M. Z{"o}llner, ``Testing and validating high level components for automated driving: Simulation framework for traffic scenarios,'' in \emph{Proceedins of the IEEE Intelligent Vehicles Symposium}, 2016, pp. 144--150.

\bibitem{NeelofarTOSEM}
N.~Neelofar and A.~Aleti, ``Identifying and explaining safety-critical scenarios for autonomous vehicles via key features,'' \emph{ACM Trans. Softw. Eng. Methodol.}, vol.~33, no.~4, Apr. 2024.

\bibitem{NeelofarICSE}
------, ``{Towards Reliable AI: Adequacy Metrics for Ensuring the Quality of System-level Testing of Autonomous Vehicles},'' in \emph{Proceedings of the IEEE/ACM 46th International Conference on Software Engineering (ICSE)}.\hskip 1em plus 0.5em minus 0.4em\relax ACM, 2024.

\bibitem{pafot}
V.~Crespo-Rodriguez, Neelofar, and A.~Aleti, ``{PAFOT: A Position-Based Approach for Finding Optimal Tests of Autonomous Vehicles},'' in \emph{Proceedings of the ACM/IEEE 5th International Conference on Automation of Software Test (AST 2024)}, ser. AST '24.\hskip 1em plus 0.5em minus 0.4em\relax ACM, 2024, p. 159–170.

\bibitem{epitester}
C.~Lu, S.~Ali, and T.~Yue, ``Epitester: Testing autonomous vehicles with epigenetic algorithm and attention mechanism,'' \emph{IEEE Transactions on Software Engineering}, pp. 1--19, 2024.

\bibitem{lu2023deepqtesttestingautonomousdriving}
\BIBentryALTinterwordspacing
C.~Lu, T.~Yue, M.~Zhang, and S.~Ali, ``{DeepQTest: Testing Autonomous Driving Systems with Reinforcement Learning and Real-world Weather Data},'' 2023. [Online]. Available: \url{https://arxiv.org/abs/2310.05170}
\BIBentrySTDinterwordspacing

\bibitem{pan2023safetyassessmentvehiclecharacteristics}
\BIBentryALTinterwordspacing
Q.~Pan, T.~Wang, P.~Arcaini, T.~Yue, and S.~Ali, ``Safety assessment of vehicle characteristics variations in autonomous driving systems,'' 2023. [Online]. Available: \url{https://arxiv.org/abs/2311.14461}
\BIBentrySTDinterwordspacing

\bibitem{KLUCK2023107225}
F.~Klück, Y.~Li, J.~Tao, and F.~Wotawa, ``An empirical comparison of combinatorial testing and search-based testing in the context of automated and autonomous driving systems,'' \emph{Information and Software Technology}, vol. 160, p. 107225, 2023.

\bibitem{10685189}
F.~Klück, D.~Sumann, and F.~Wotawa, ``Utilizing genetic algorithms for generating critical scenarios for testing autonomous driving functions,'' in \emph{2024 IEEE International Conference on Artificial Intelligence Testing (AITest)}, 2024, pp. 73--80.

\bibitem{HUMENIUK2023102990}
D.~Humeniuk, F.~Khomh, and G.~Antoniol, ``Ambiegen: A search-based framework for autonomous systems testingimage 1,'' \emph{Science of Computer Programming}, vol. 230, p. 102990, 2023.

\bibitem{10.1145/3650105.3652296}
J.~Wu, C.~Lu, A.~Arrieta, T.~Yue, and S.~Ali, ``Reality bites: Assessing the realism of driving scenarios with large language models,'' in \emph{Proceedings of the IEEE/ACM 2024 First International Conference on AI Foundation Models and Software Engineering (FORGE)}.\hskip 1em plus 0.5em minus 0.4em\relax ACM, 2024, p. 40–51.

\bibitem{ARCAINI2024103171}
P.~Arcaini and A.~Cetinkaya, ``Crag – a combinatorial testing-based generator of road geometries for ads testing,'' \emph{Science of Computer Programming}, vol. 238, p. 103171, 2024.

\bibitem{10.1145/3550270}
T.~Laurent, S.~Klikovits, P.~Arcaini, F.~Ishikawa, and A.~Ventresque, ``Parameter coverage for testing of autonomous driving systems under uncertainty,'' \emph{ACM Trans. Softw. Eng. Methodol.}, vol.~32, no.~3, Apr. 2023.

\bibitem{10.1007/978-3-031-49269-3_14}
F.~Khan, H.~Anwar, and D.~Pfahl, ``Simulation-based safety testing of automated driving systems,'' in \emph{Product-Focused Software Process Improvement}.\hskip 1em plus 0.5em minus 0.4em\relax Springer, 2024, pp. 133--138.

\bibitem{10.1007/978-3-031-49266-2_6}
------, ``A process for scenario prioritization and selection in simulation-based safety testing of automated driving systems,'' in \emph{Product-Focused Software Process Improvement}.\hskip 1em plus 0.5em minus 0.4em\relax Springer, 2024, pp. 89--99.

\bibitem{2025-Baresi-ICSE}
L.~Baresi, D.~Y.~X. Hu, A.~Stocco, and P.~Tonella, ``Efficient domain augmentation for autonomous driving testing using diffusion models,'' in \emph{Proceedings of the IEEE 47th International Conference on Software Engineering}, ser. ICSE '25.\hskip 1em plus 0.5em minus 0.4em\relax IEEE, 2025.

\bibitem{PDL}
Q.~Ali, O.~Riganelli, and L.~Mariani, ``Testing in the evolving world of dl systems: Insights from python github projects,'' in \emph{2024 IEEE 24th International Conference on Software Quality, Reliability and Security (QRS)}, 2024, pp. 25--35.

\bibitem{sensodat}
\BIBentryALTinterwordspacing
C.~Birchler, C.~Rohrbach, T.~Kehrer, and S.~Panichella, ``Sensodat: Simulation-based sensor dataset of self-driving cars,'' in \emph{Proceedings of the 21st International Conference on Mining Software Repositories}, ser. MSR '24.\hskip 1em plus 0.5em minus 0.4em\relax New York, NY, USA: Association for Computing Machinery, 2024, p. 510–514. [Online]. Available: \url{https://doi.org/10.1145/3643991.3644891}
\BIBentrySTDinterwordspacing

\bibitem{deepscenario}
C.~Lu, T.~Yue, and S.~Ali, ``Deepscenario: An open driving scenario dataset for autonomous driving system testing,'' in \emph{2023 IEEE/ACM 20th International Conference on Mining Software Repositories (MSR)}, 2023, pp. 52--56.

\bibitem{sctrans}
\BIBentryALTinterwordspacing
J.~Dai, B.~Gao, M.~Luo, Z.~Huang, Z.~Li, Y.~Zhang, and M.~Yang, ``Sctrans: Constructing a large public scenario dataset for simulation testing of autonomous driving systems,'' in \emph{Proceedings of the IEEE/ACM 46th International Conference on Software Engineering}, ser. ICSE '24.\hskip 1em plus 0.5em minus 0.4em\relax New York, NY, USA: Association for Computing Machinery, 2024. [Online]. Available: \url{https://doi.org/10.1145/3597503.3623350}
\BIBentrySTDinterwordspacing

\bibitem{borg}
M.~Borg, R.~B. Abdessalem, S.~Nejati, F.-X. Jegeden, and D.~Shin, ``Digital twins are not monozygotic--cross-replicating adas testing in two industry-grade automotive simulators,'' in \emph{ICST '21}.\hskip 1em plus 0.5em minus 0.4em\relax IEEE, 2021.

\bibitem{AminiFlaky2024}
M.~H. Amini, S.~Naseri, and S.~Nejati, ``Evaluating the impact of flaky simulators on testing autonomous driving systems,'' \emph{Empirical Softw. Engg.}, vol.~29, no.~2, feb 2024.

\bibitem{beamng}
BeamNG.tech, ``{BeamNG GmbH},'' \url{ https://beamng.tech/}, 2024, online; accessed 2024-11-19.

\bibitem{udacity_sim}
\BIBentryALTinterwordspacing
Udacity, ``Self-driving car simulator,'' 2023, accessed: 2024-11-04. [Online]. Available: \url{https://github.com/udacity/self-driving-car-sim}
\BIBentrySTDinterwordspacing

\bibitem{dupuis2010opendrive}
M.~Dupuis, M.~Strobl, and H.~Grezlikowski, ``{OpenDRIVE 2010 and beyond – status and future of the de facto standard for the description of road networks},'' in \emph{Proceeding of the Driving Simulation Conference Europe}, 2010, pp. 231--242.

\bibitem{Catmull1974ACO}
\BIBentryALTinterwordspacing
E.~E. Catmull and R.~Rom, ``A class of local interpolating splines,'' \emph{Computer Aided Geometric Design}, pp. 317--326, 1974. [Online]. Available: \url{https://api.semanticscholar.org/CorpusID:118383557}
\BIBentrySTDinterwordspacing

\bibitem{nhtsa}
U.~D. of~Transportation, ``A framework for automated driving system testable cases and scenarios,'' \url{https://rosap.ntl.bts.gov/view/dot/38824/dot_38824_DS1.pdf}, 2018.

\bibitem{nvidia-dave2}
M.~Bojarski, D.~Del~Testa, D.~Dworakowski, B.~Firner, B.~Flepp, P.~Goyal, L.~D. Jackel, M.~Monfort, U.~Muller, J.~Zhang, X.~Zhang, J.~Zhao, and K.~Zieba, ``End to end learning for self-driving cars.'' \emph{CoRR}, vol. abs/1604.07316, 2016.

\bibitem{udacity-simulator}
{Udacity}, ``Udacity self-driving car simulator,'' \url{https://github.com/udacity/self-driving-car-sim}, 2021, accessed: [2024-01-15].

\bibitem{asam_opendrive}
{Association for Standardization of Automation and Measuring Systems (ASAM)}, ``{ASAM OpenDRIVE®},'' \url{https://www.asam.net/standards/detail/opendrive/}, 2023, accessed: 2024-11-04.

\bibitem{baass1984clothoid}
K.~Baass, ``The use of clothoid templates in highway design,'' in \emph{Transportation Forum}, vol.~1, 1984, pp. 47--52.

\bibitem{isprs}
\BIBentryALTinterwordspacing
B.~Schwab and T.~H. Kolbe, ``Validation of parametric opendrive road space models,'' \emph{ISPRS Annals of the Photogrammetry, Remote Sensing and Spatial Information Sciences}, vol. X-4/W2-2022, pp. 257--264, 2022. [Online]. Available: \url{https://isprs-annals.copernicus.org/articles/X-4-W2-2022/257/2022/}
\BIBentrySTDinterwordspacing

\bibitem{klikovits2022doesroaddiversityreally}
\BIBentryALTinterwordspacing
S.~Klikovits, V.~Riccio, E.~Castellano, A.~Cetinkaya, A.~Gambi, and P.~Arcaini, ``Does road diversity really matter in testing automated driving systems? -- a registered report,'' 2022. [Online]. Available: \url{https://arxiv.org/abs/2209.05947}
\BIBentrySTDinterwordspacing

\bibitem{catmull-rom}
\BIBentryALTinterwordspacing
S.~Documentation, ``Catmull-rom splines,'' 2024, accessed: 2024-11-19. [Online]. Available: \url{https://splines.readthedocs.io/en/latest/euclidean/catmull-rom.html}
\BIBentrySTDinterwordspacing

\bibitem{opendrive2catmullrom}
\BIBentryALTinterwordspacing
Q.~Ali, ``Opencat,'' 2024, accessed: 2024-11-19. [Online]. Available: \url{https://github.com/lakhanqurban/OpenCat}
\BIBentrySTDinterwordspacing

\bibitem{libOpenDRIVE}
\BIBentryALTinterwordspacing
M.~Pagel \emph{et~al.}, ``libopendrive,'' 2024, accessed: 2024-11-04. [Online]. Available: \url{https://github.com/pageldev/libOpenDRIVE}
\BIBentrySTDinterwordspacing

\bibitem{pagel_odrviewer}
S.~Pagel, ``Opendrive viewer,'' \url{https://sebastian-pagel.net/odrviewer/}, accessed: 2024-11-04.

\bibitem{biagiola2024}
\BIBentryALTinterwordspacing
M.~Biagiola, A.~Stocco, V.~Riccio, and P.~Tonella, ``Two is better than one: digital siblings to improve autonomous driving testing,'' \emph{Empirical Software Engineering}, vol.~29, p.~72, 2024. [Online]. Available: \url{https://doi.org/10.1007/s10664-024-10458-4}
\BIBentrySTDinterwordspacing

\bibitem{gambi2019automatically}
A.~Gambi, M.~Mueller, and G.~Fraser, ``Automatically testing self-driving cars with search-based procedural content generation,'' in \emph{Proceedings of the ACM 28th SIGSOFT International Symposium on Software Testing and Analysis}, 2019, pp. 318--328.

\end{thebibliography}

\end{document}